\documentclass[conference]{IEEEtran}

\usepackage{cite}
\usepackage{graphicx}
\usepackage{epstopdf}
\usepackage{enumerate}
\usepackage{booktabs}
\usepackage{multirow}
\usepackage{tabularx}
\usepackage{array}
\usepackage{pifont}
\usepackage{amsmath}
\usepackage{color}
\usepackage{amssymb}
\usepackage{amsmath}
\usepackage{bm}
\usepackage{tabulary}
\usepackage{flushend}
\IEEEoverridecommandlockouts
\allowdisplaybreaks
\begin{document}
\newcommand{\minitab}[2][l]{\begin{tabular}{#1}#2\end{tabular}} 
\renewcommand\arraystretch{1.2}
% paper title
\title{SFCSD: A Self-Feedback Correction System for DNS Based on Active and Passive Measurement}

\author{
\IEEEauthorblockN{Caiyun Huang\IEEEauthorrefmark{1}\IEEEauthorrefmark{2}, Peng Zhang\IEEEauthorrefmark{2}, Junpeng Liu\IEEEauthorrefmark{2}, Yong Sun\IEEEauthorrefmark{2}, Xueqiang Zou\IEEEauthorrefmark{2}\IEEEauthorrefmark{3}} 
  
\IEEEauthorblockA{\IEEEauthorrefmark{1}School of Cyber Security University of Chinese Academy of Sciences, Beijing, China} 
\IEEEauthorblockA{\IEEEauthorrefmark{2}Institute of Information Engineering, Chinese Academy of Sciences, Beijing, China}
\IEEEauthorblockA{\IEEEauthorrefmark{3}National Computer Network Emergency Response and Coordination Center, Beijing, China}
Emails: \{huangcaiyun, pengzhang, liujunpeng, sunyong, zouxueqiang\}@iie.ac.cn 
}

% make the title area
\maketitle
\begin{abstract}
Domain Name System (DNS), one of the important infrastructure in the Internet, was vulnerable to attacks, for the DNS designer didn't take security issues into consideration at the beginning. The defects of DNS may lead to users' failure of access to the websites, what's worse, users might suffer a huge economic loss.

In order to correct the DNS wrong resource records, we propose a Self-Feedback Correction System for DNS (SFCSD), which can find and track a large number of common websites' domain name and IP address correct correspondences to provide users with a real-time auto-updated correct (IP, Domain) binary tuple list. By matching specific strings with SSL, DNS and HTTP traffic passively, filtering with the CDN CNAME and non-homepage URL feature strings, verifying with webpage fingerprint algorithm, SFCSD obtains a large number of highly possibly correct IP addresses to make an active manual correction in the end. Its self-feedback mechanism can expand search range and improve performance.

Experiments show that, SFCSD can achieve 94.3\% precision and 93.07\% recall rate with the optimal threshold selection in the test dataset. It has 8Gbps processing speed stand-alone to find almost 1000 possibly correct (IP, Domain) per day for the each specific string and to correct almost 200.
\end{abstract}
\bigskip

\IEEEpeerreviewmaketitle

\section{Introduction}

Domain Name System (DNS) is one of the important infrastructure in the Internet, providing correspondence between domain name and IP address for almost all kinds of the upper applications. As the DNS designer didn't consider the security issues at the beginning \cite{Mockapetris1987}, many attackers are targeting DNS. Once the DNS Resource Records (RRs) are incorrect, they may cause a large scale of network paralysis \cite{Wikipedia2017}.

\begin{figure}
\centering
\includegraphics[height=6cm]{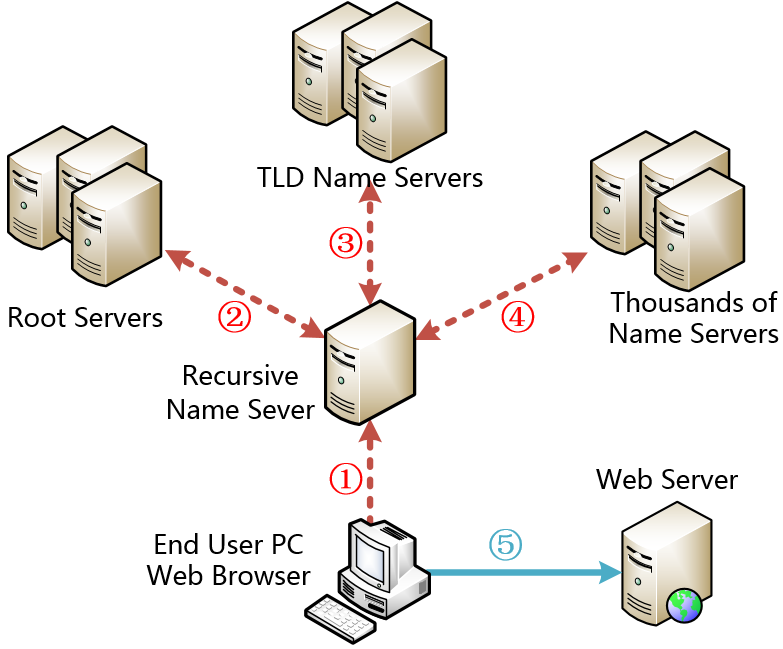}
\caption{After a user press Enter.}
\label{fig:introduction}
\end{figure}

In Figure \ref{fig:introduction}, a user types a domain name in the browser and presses Enter at first. Then, as is shown in step 1 to step 4, through a series of DNS requests and responses among the personal computer and each layer domain name server in the back, the browser gets one or several IP addresses which have access to the websites successfully in the last step 5. Under normal circumstances, the browser can get the correct IP addresses to access the correct websites.

However, except the non-malicious server configuration error \cite{Gao2016}, in the process represented by the dashed arrows, there are many causes for incorrect DNS RRs, such as DNS spoofing \cite{hanley2000dns}, DNS cache poisoning \cite{Son2010}. Attackers can utilize DNS spoofing by pretending to be the domain name server to return a wrong IP address to the browser \cite{yan2006detection}, or use DNS cache poisoning to inject illegal domain name into DNS server's cache \cite{Son2010}, or even more, they can intercept DNS requests to make them lose their responses within a range of network to DNS hijacking \cite{larson2005dns}.

There are already a lot of solutions for the above-mentioned attacks. In this paper, the main idea is to find correct (IP, Domain), not to prevent the attacks, but to avoid them. We propose a fast, self-feedback system to find and track a large number of common websites' domain name and IP address correct correspondences with a real-time dynamic auto-updated mechanism by using passive measurement with SSL, DNS and HTTP traffic, filtering with the CDN CNAME and non-homepage URL feature strings, and verifying with webpage fingerprint algorithm and active functions.

In general, the major contributions of this paper are as follows:

\begin{itemize}
\item A fast and self-feedback system is proposed which can continuously track and find the correct correspondence between the specific domain name and IP address online;
\item Summary 8 CDN service provider's CNAME feature strings to filter out the CDN server IP and 8 non-homepage URL characteristics to filter out the non-homepage server IP, which reduces the computing pressure of online correction and improves the system precision;
\item Utilize both active and passive measurement to recommend the passive results online with a certain error rate and the active 100\% correct results offline with a little delay.
\end{itemize}

This paper is arranged as follows: Section \ref{sec:relatedwork} introduces the previous related work. Section \ref{sec:framework} is about the system architecture and the four core modules briefly. Section \ref{sec:methoddesign} describes the key technologies applied in each core module in detail. Section \ref{sec:experiments} evaluates the effect of this system through the experiments. Section \ref{sec:discussion} discusses the limitation of the system and the future work. Finally, the conclusion is contained in Section \ref{sec:conclusion}.

\section{Related Work}
\label{sec:relatedwork}
As mentioned above, the main cause for the incorrect DNS RRs is that there are a variety of attacks to DNS. The current research hotspot could be divided into two types of ideas, one is using the DNS protocol extension fields plus encryption technology to guarantee the security of communication. Zou et al. \cite{zou2016survey} introduced 8 kinds of DNS security architecture which is based on the current DNS protocol to expand, such as DNSSEC \cite{larson2005dns}, DNSCurve \cite{bernstein2009dnscurve}, WSECDNS \cite{perdisci2009wsec}, DDNS \cite{Cox2002} and CoDoNS \cite{park2004codns}, all of them are the conclusion solution to overcome weakness in the DNS protocol.

Another is for one specific type of DNS attacks to design protection scheme without changing protocol. Regarding the DNS spoofing, Alqahtani et al. \cite{alqahtani2013tcp} proposed to use IP address based authentication instead of using domain name based. For the DNS cache poisoning, \cite{park2004codns,hmood2015adaptive} design a methodology called Adaptive-Cache of DNS by caching mechanism and backward compatible with the current standards of DNS.

These solutions mainly aim to detect and defend the causing factors, but the main idea of this paper is different from them. SFCSD don't defend against the attack directly, however, in case of the existence of these attacks, it analyzes the DNS RRs and filter out the wrong one which can't directly access the websites correctly. From the point of view of attackers, they fail in the last step of the attacks. From the user accessing to the common websites, they directly bypass the possible attacks in the first step.

The idea of Jain et al. \cite{jain2016novel} is similar to ours, both can build an auto-updated whitelist, but the difference is that SFCSD find the correct and available correspondence of domain name and IP which is not only correct in resolution, but also can access correct common websites. While they use the third-party Google public domain name server to resolute the domain name, Kuhrer et al. \cite{kuhrer2015going} find that the open domain name server is increasingly vulnerable to attacks, so these third-party may also provide a wrong DNS RRs.

\section{Framework}
\label{sec:framework}
\begin{figure*}
\centering
\includegraphics[height=8cm]{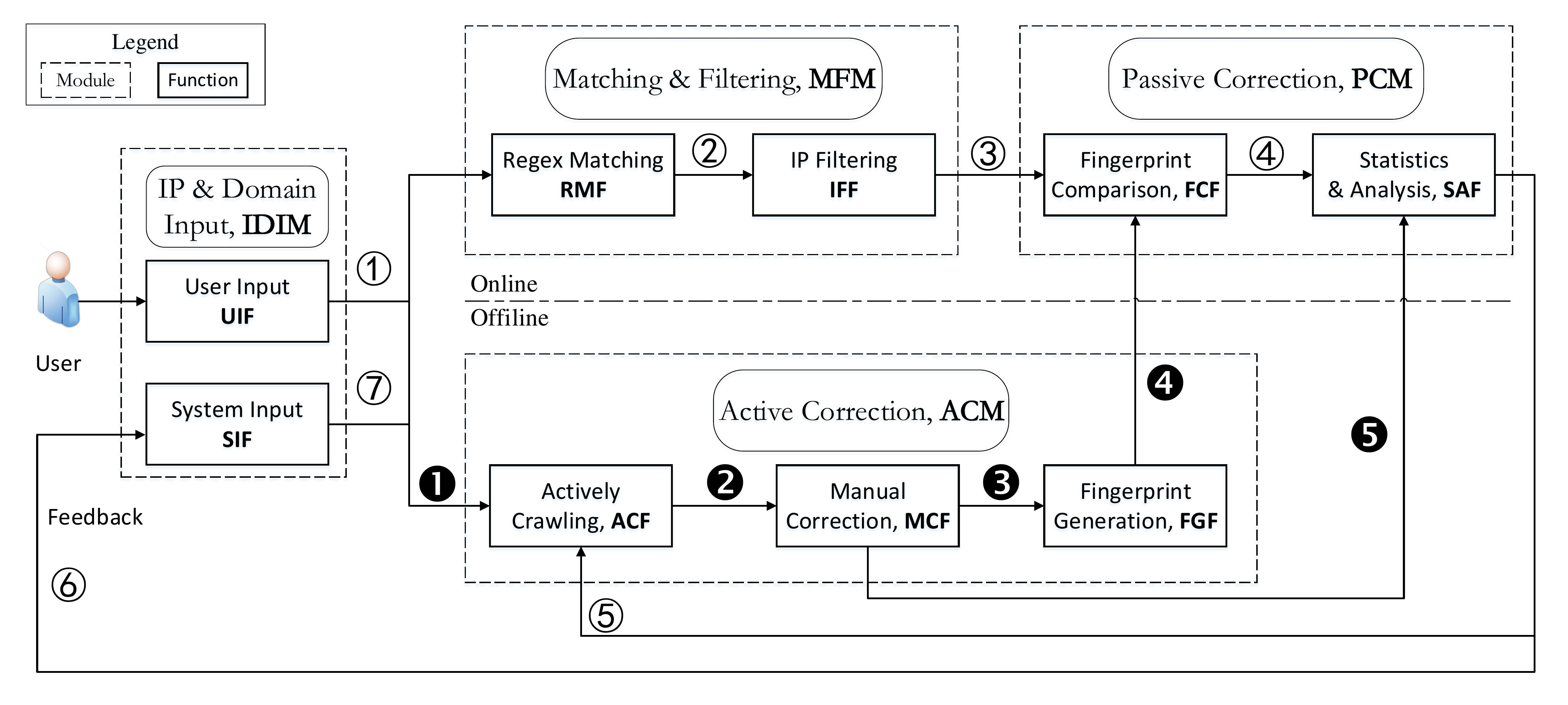}
\caption{The overview of the self-feedback system for DNS based on active and passive measurement.}
\label{fig:framework-all}
\end{figure*}

As shown in Figure \ref{fig:framework-all}, SFCSD can be divided into four core modules, IP \& Domain Input Module (IDIM), Matching \& Filtering Module (MFM), Passive Correction Module (PCM) and Active Correction Module (ACM).

IDIM receives specific IP \& Domain strings from the user and system itself, then inputs them to MFM and ACM. MFM resolves the passive traffic to match the IP \& Domain strings and outputs initially suspicious (IP, Domain) to PCM. PCM combines with ACM to correct the initial outputs and sends recommended (IP, Domain) to IDIM as the feedback. Each Module has its own Functions and will be described here.

\subsection{IP \& Domain Input Module}
As displayed in Figure \ref{fig:framework-all}, IDIM consists of two Functions, System Input Function (SIF) and User Input Function (UIF). SIF is to receive the specific strings from SFCSD itself, which is automatically generated by PCM. UIF is user-oriented to obtain users' interests. The specific strings are classified into two types, Domain string and IP string. Both of them can be input by the user, while the IP string can also be auto-generated. IDIM sends the specific strings to MFM and ACM as the first filter conditions to narrow the focus range, and its auto-generated IP string is the basis of the self-feedback mechanism to expand search range and improve the whole system's effects.

\subsection{Matching \& Filtering Module}
As Figure \ref{fig:framework-all} indicates, MFM contains two Functions, Regex Matching Function (RMF) and IP Filtering Function (IFF). RMF extracts all relevant IP, Domain, CNAME and webpage contents which match the specific strings from IDIM by resolving rapidly large-scale SSL, DNS and HTTP traffic based on previous work \cite{liu2014towards}. Its main function is to narrow the focus range and reduce unnecessary overload. IFF receives RMF's outputs to filter out the CDN server IP and non-homepage URL by some feature strings, and sends the suspicious (IP, Domain) to PCM for further analyses. Its main function is, under the conditions of guaranteeing the online processing speed and the system recall rate, to reduce PCM's processing pressure as much as possible.

\subsection{Passive Correction Module}
As suggested in Figure \ref{fig:framework-all}, PCM is made up of two Functions, Fingerprint Comparison Function (FCF) and Statistics \& Analysis Function (SAF). FCF generates webpage fingerprints online with the webpage contents sent by MFM, which are all possible homepages filtered by IFF, and compares them with the standard homepage websites fingerprints generated by ACM to calculate the similarity of fingerprints. Its main function is to correct the suspicious (IP, Domain) which appears in HTTP by utilizing webpage fingerprint algorithm.

SAF analyzes the outputs from MFM first, if the correctness possibility of (IP, Domain) is more than a certain threshold ${T_1}$ such as 98\%, it will provide them for users and send them to IDIM directly as a feedback. While if the (IP, Domain) are very suspicious but not sure, namely the possibility is smaller than ${T_1}$ but larger than ${T_2}$ such as 90\%, it will send them to ACM as recommended for further correction, and receives ACM correction results to generate the IP string too.

\subsection{Active Correction Module}
As Figure \ref{fig:framework-all} demonstrates, ACM constitutes three Functions, Actively Crawling Function (ACF), Manual Correction Function (MCF) and Fingerprint Generation Function (FGF). ACF receives the specific strings from IDIM and the recommended (IP, Domain) from PCM to crawl relevant data by some active methods. However, due to some wrong cases, SFCSD needs MCF to correct by people. MCF's main function is picking out the 100\% correct websites homepages to FGF, correcting the recommended (IP, Domain) and outputting the correction results to PCM and IDIM. FGF receives the completely correct websites homepages and generates the websites fingerprints for FCF, which is to improve online process speed.

%IP string auto-generation
\section{Method Design}
\label{sec:methoddesign}
% The following describes the core technologies of the above each module, including SIF \& SAF in IDIM \& PCM, IFF in MFM, FCF \& FGF in PCM \& ACM, ACF \& MCF in ACM. They achieve the self-feedback mechanism, the IP filtering, the passive correction and the active correction functions respectively.
SFCSD is a self-feedback correction system for DNS based on active and passive measurement. This section describes the core technologies about the self-feedback mechanism, the passive correction and the active correction respectively as follows.

\subsection{Self-feedback Mechanism}
All functions work together to achieve the self-feedback mechanism. While the most important is SIF in IDIM and SAF in PCM.

\subsubsection{IP String Auto-Generation}
As mentioned above, SIF is a basis of self-feedback mechanism, its goal is to obtain related traffic at most and narrow the focus range. As the system continues to run, PCM will automatically generate the IP string to IDIM, then IDIM sends these to MFM and ACM in order to find the new (IP, domain) tuples.

At the initial stage of the system, there is no auto-generated IP string, all of IDIM outputs are from users. When users input a Domain string like ``example.com", MFM resolves the DNS, SSL and HTTP traffic and extracts all relevant server IP matching this Domain string, by PCM and ACM, the system will generate the IP string to IDIM. When the later traffic come, MFM will matching this IP in the traffic too, if the (IP, Domain) always appears together in PCM recommended results, PCM regards them as the corrected results to users. The more frequently a (IP, Domain) appears, the more likely it is correct.

\subsubsection{Possibility Calculating Method}
The main function of SAF is to calculate the possibility of the correctness of all (IP, Domain) found in traffic. The possibility calculating method is the theoretical basis for the self-feedback mechanism as follows.

The initial possibility of the (IP, Domain) first found in the passive traffic is $0\%$, after matched by RMF, each (IP, Domain) matched with the specific strings has a possibility depending on the protocol which the (IP, Domain) come from. The possibility of SSL is denoted as $s\%$, DNS is $d\%$ and HTTP is $h\%$, because of the different rank of protocol security, $s\%>h\%>d\%$. If a (IP, Domain) appears in two or three protocols at a short time window such as 5 minutes, its possibility will increase as Equation (\ref{eq:ssldns}) shows. In a similar procedure, $(SSL\&HTTP)=(s+h)\%-sh\%$, $(DNS\& HTTP) = (d + h)\%  - dh\%$, $(SSL\& DNS\& HTTP)=(s + d + h)\%  - sd\%  - sh\%  - dh\%  + sdh\%$.

\begin{equation}
\begin{split}
(SSL\& DNS) & = s\%  + (1 - s\% ) \times d\% \\
 &= d\%  + (1 - d\% ) \times s\% \\
 &= s\%  + d\%  - s\%  \times d\% \\
 &= (s + d)\%  - sd\% 
\end{split}
\label{eq:ssldns}
\end{equation}

And then IFF filters out the CDN server IP and non-homepage URL relevant IP, which means to set these (IP, Domain)'s possibilities as $0\%$. 
PCM gets all (IP, Domain) whose possibility more than $0\%$, and if the (IP, Domain) come from HTTP, there is a further correction. By computing the similarity of fingerprints denoted as $sim\%$, the more similar fingerprints are, the more likely the (IP, Domain) is correctly, and the possibility is calculated in the same way.

SAM obtains the suspicious (IP, Domain) and computes their possibilities. If the possibility meet the certain threshold ${T_1}$, SFCSD sends the passive results online with a certain error rate to users. If the possibility is smaller than ${T_1}$ but larger than ${T_2}$, SFCSD will sends these (IP, Domain) to ACM.

Since some web servers don't response the IP requests, there are two kinds of results by ACM, 100\% correct and still uncertain. ACM sends these results back to SAF, and then SAF issues the 100\% correct (IP, Domain) to users with a little delay. All these judgment results by ACM or SAF are available in a period of time such as one day.

Except above two operations, SAM will send all suspicious (IP, Domain) to IDIM as a feedback, SIF receives these (IP, Domain) with the user input and transmit all to RMF.

So far, the possibility of the correctness of (IP, Domain) which is first found in traffic is calculated over. Given the IP string auto-generation as the feedback, SAF will calculate the (IP, Domain) occurrence times and last time, if the (IP, Domain) don't appear one day, its possibility will decrease $dec\%$, in contrast, if the (IP, Domain) appear one day, its possibility will increase $inc\%$ until the threshold ${T_1}$.

\subsection{Passive Correction}
The passive correction is a cooperation of IFF in MFM, FGF in ACM and FCF in PCM. It compares the homepage fingerprints generated by FGF with the online webpage filtered by IFF and fingerprints generated by FCF for their similarity to correct the (IP, Domain).

Its main idea is through denoting a homepage as its fingerprint and denoting the similarity between homepage and webpage as the hamming distance between the two fingerprints to find the webpage from MFM similar to the corresponding websites' homepage, which means the server IP in this HTTP packet has a great possibility of being the website's web server IP.

\subsubsection{IP Filtering}
IFF is the core technology of MFM. The main goal is to filter out the Web server IP addresses that completely can't directly access a specific website homepage in all traffic matching the specific string to reduce the processing pressure of PCM. On the basis of ensuring the system recall rate, IFF is to improve the system processing speed as much as possible, by filtering CDN server IP in DNS and non-homepages in HTTP.

\paragraph{CDN Filter}
CDN Filter mainly filters out the IP address of the CDN server from all server IP addresses found in DNS. Since currently many websites will use CDN to accelerate the webpages loading speed, and users visit the CDN server first, then the CDN server accesses the webpage from the source websites to back to users. So there are many CDN server IP addresses found from the DNS traffic resolution, which aren't the real web server IP address. The key of the CDN filter is that almost all the CDN service providers will have an obvious feature string in CNAME record, as Table \ref{tab:cdn_feature} shows. The filtration steps are as follows:

\begin{enumerate}
\item[\ding{192}] Extract the CNAME requests from the DNS;
\item[\ding{193}] If the extract CNAME contains the feature strings in Table \ref{tab:cdn_feature}, add the server IP to the CDN server IP list, otherwise do nothing;
\item[\ding{194}] Output all suspicious server IP which aren't in CDN server IP list.
\end{enumerate}

\begin{table}
\caption{CDN Filter Strings}
\centering
\begin{tabular}{|c|c|c|}
\hline
\bfseries No. &\bfseries CDN Service Provider &\bfseries Feature Strings in CNAME\\ 
\hline
\multirow{4}*{1} &\multirow{4}*{Akamai} & akamai \\ %\cline{3-3}
& & akadns \\ %\cline{3-3}
& & edgesuite \\ %\cline{3-3}
& & edgekey \\ %\cline{3-3}
\hline
\multirow{2}*{2} &\multirow{2}*{Amazon Cloudfront} & cloudfront \\ %\cline{3-3}
& & amazon \\ 
\hline
\multirow{1}*{3} & ChinaCache & chinacache \\
\hline
\multirow{1}*{4} & CloudFlare & cloudflare \\
\hline
\multirow{1}*{5} & EdgeCast & edgecast \\
\hline
\multirow{1}*{6} & Fastly & fastly \\
\hline
\multirow{1}*{7} & Incapsula & incap \\
\hline
\multirow{2}*{8} & \multirow{2}*{Tencent} & cdntip \\ %\cline{3-3}
& & dnsv1 \\
\hline
\multirow{3}*{9} & \multirow{3}*{Other} & passvpn \\ %\cline{3-3}
& & a-msedge1 \\ %\cline{3-3}
& & edge \\ %\cline{3-3}
\hline
\end{tabular}
\label{tab:cdn_feature}
\end{table}

\paragraph{URL Filter}
URL Filter is mainly filters out all non-homepages in HTTP, because FGF considers the homepage as a symbol to represent the website, so it only generates the fingerprint for the homepage. If URL Filter can ensure that the two comparing pages are homepages, which can improve speed and effect of FCF. The filtration steps are as follows:
\begin{enumerate}
\item[\ding{192}] Extract the URL of the HTTP packets;
\item[\ding{193}] Filter the URL by the simple rules in Table \ref{tab:url_feature};
\item[\ding{194}] In order to ensure the recall rate, only when the URL matching the non-homepages URL features, out the server IP and HTTP packets;
\end{enumerate}

The key of URL Filter is to find useful homepage URL characteristics. Due to the URL redirection and URL rewrites in the web server, not all URLs with only domain can access the homepage, and there are some other URL patterns can access the homepage either. Thus, the 8 URL patterns are proposed in Table \ref{tab:url_feature}, T indicates the URL with that pattern is a homepage, and F indicates not. We realize these patterns by regular expression which can reduce the delay online as much as possible. 

% T/F T、F
\begin{table}
\caption{URL Filter Rules}
\centering
\begin{tabular}{|c|c|c|c|}
\hline
\bfseries No. &\bfseries URL Patterns &\bfseries Examples &\bfseries T/F \\
\hline
\multirow{2}*{1} & \multirow{2}*{\minitab[c]{With only domain}} & \multirow{2}*{\minitab[c]{www.example.com}} & \multirow{2}*{\minitab[c]{T}} \\
  & & & \\
\hline
\multirow{2}*{2} & \multirow{2}*{\minitab[c]{With domain less \\than 5 characters}} & \multirow{2}*{\minitab[c]{www.example.com\\ /\#doj}} & \multirow{2}*{\minitab[c]{T}} \\
  & & & \\
\hline
\multirow{2}*{3} & \multirow{2}*{\minitab[c]{Containing strings \\ like ``index.html"}} & \multirow{2}*{\minitab[c]{www.example.com \\ /.../index.html}} & \multirow{2}*{\minitab[c]{T}} \\
  & & & \\
\hline
\multirow{2}*{4} & \multirow{2}*{\minitab[c]{Containing non-text \\ suffix like ``.exe"}} & \multirow{2}*{\minitab[c]{www.example.com \\ /.../download.exe}} & \multirow{2}*{\minitab[c]{F}} \\
  & & & \\
\hline
\multirow{2}*{5} & \multirow{2}*{\minitab[c]{With several numbers \\followed by suffix}} & \multirow{2}*{\minitab[c]{www.example.com \\ /.../n47827153.html}} & \multirow{2}*{\minitab[c]{F}} \\
  & & & \\
\hline
\multirow{2}*{6} & \multirow{2}*{\minitab[c]{Containing strings \\ like ``redirect"}} & \multirow{2}*{\minitab[c]{www.example.com \\ /.../redirect...}} & \multirow{2}*{\minitab[c]{F}} \\
  & & & \\
\hline
\multirow{2}*{7} & \multirow{2}*{\minitab[c]{With over \\ 2 parameters}} & \multirow{2}*{\minitab[c]{www.example.com \\ /.../?a=1\&c=2\&c=3}} & \multirow{2}*{\minitab[c]{F}} \\
  & & & \\
\hline
\multirow{2}*{8} &  \multirow{2}*{\minitab[c]{With over \\ 50 characters}} & \multirow{2}*{\minitab[c]{www.example.com \\ /.../.../...}} & \multirow{2}*{\minitab[c]{F}} \\
  & & & \\
\hline
\end{tabular}
\label{tab:url_feature}
\end{table}

% \newgeometry{left=0.7in,right=0.7in}

\subsubsection{Fingerprint Generation}
The Fingerprint Generation algorithm is SimHash \cite{sadowski2007simhash}, which is used to check massive text duplicates by Google. Through a series of operations, a document can be transferred into a 64-bit string, namely the fingerprint, and then by computing the hamming distance between the two fingerprints to determine whether the two documents are similar, if the distance is less than a certain threshold N, they are.

In short, SimHash steps are as follows:
\begin{enumerate}
\item[\ding{192}] Participle: participle the source text into some characteristic words, and remove the noise words to build a standard word sequence and weight for each word;
\item[\ding{193}] Hash: hash each word into a string of numbers to reduce the document dimension;
\item[\ding{194}] Weight: according to the word weight, process the string of numbers generated in step 2 to form a weighted string;
\item[\ding{195}] Unite: for each word, accumulate all the string of numbers to one;
\item[\ding{196}] Reduce dimension: change the string of numbers from the Step 2 into binary string, namely the fingerprint.
\end{enumerate}

\subsubsection{Fingerprint Matching}
Through the above steps, FGF generates the specific websites' fingerprints into database. FCF obtains the webpage contents which are filtered by URL and feature string, and then generates the webpage fingerprints online and calculates the distance between the two webpage Fingerprints. The smaller the distance is, the more possible to find the web server IP address is.

\subsection{Active Correction}
\subsubsection{Actively Crawling}
In addition to obtaining the required IP address corresponding of the domain, it is necessary to obtain the homepage of the specific websites and generate the websites fingerprints. For the former, ACF utilizes the following four methods to obtain, and for the latter, ACF uses spiders like Wget \cite{Scrivano2017} to crawl automatically.

\begin{itemize}
\item DNS server resolution: By simulating normal users accessing a website to obtain the DNS response from the domain name server and find the web server IP;
\item Third-party database: By collecting the required data from the third-party IP database, such as ARIN \cite{ARIN2017}, RIPE \cite{Ripencc2017}, APNIC \cite{APNIC2017} and IPIP \cite{IPIP2017}, etc;
\item Ping service online: By using the ping tools provided by the websites in Table \ref{tab:test_websites} to get the specific domain corresponding IP;
\item Hosts tools: By tracking and crawling the popular hosts tools in Table \ref{tab:hosts_tools} periodically to obtain the web server IP.
\end{itemize}

\begin{table}
\caption{Popular Ping Service Online}
\centering
\begin{tabular}{cll}
\hline
\bfseries No. &\bfseries Name &\bfseries Websites \\
\hline
1&ping.eu & http://ping.eu/ping/ \\
2&CA App Synthetic Monitor & https://asm.ca.com/en/ping.php \\
3&HostTracker & http://host-tracker.com \\
4&kakawang & http://www.webkaka.com \\
5&Internet Supervision & http://internetsupervision.com \\
6&yperSpin & http://www.hyperspin.com \\
7&Alertra & http://www.alertra.com \\
\hline
\end{tabular}
\label{tab:test_websites}
\end{table}

\begin{table}
\caption{Popular Hosts Tools}
\centering
\begin{tabular}{cll}
\hline
\bfseries No. &\bfseries Name &\bfseries Websites \\
\hline
1&huhamhire-hosts & http://code.google.com/p/huhamhire-hosts \\
2&Smarthosts & http://code.google.com/p/smarthosts \\
3&HostsX & http://code.google.com/p/hostsx \\
\hline
\end{tabular}
\label{tab:hosts_tools}
\end{table}

\subsubsection{Manual Correction}
Since the standard of correct (IP, Domain) is to directly access the specific websites correctly, while the data obtained by ACF contains some wrong cases, MCF has further correction as follows to make sure the homepages and (IP, Domain) are 100\% correct:

\begin{enumerate}
\item[\ding{192}] Modify the local HOSTS file with the (IP, Domain) to be corrected;
\item[\ding{193}] Simulate a browser accessing the websites by Webdriver \cite{Selenium2017};
\item[\ding{194}] Judge the correctness of (IP, Domain) based on the response;
\item[\ding{195}] If the program can't judge definitely, it's time to people help.
\end{enumerate}

\section{Evaluation}
\label{sec:experiments}
This section will evaluate the system effects about the recall, accuracy, precision rate and online processing performance.

\subsection{Dataset}
We selected the top 50 Chinese websites from the Alexa \cite{Alexa2017} rank as users' inputs, given the workload of manual labeling, we only acquired running logs from the online testing program in the CSTNET \cite{Wikipedia} within 2 hours. The overview of the dataset is in Table \ref{tab:dataset_outline}. 

Each packet matching the specific strings will generate a log which contains the packet information, including protocol, matching string, server IP, and host name for SSL, domain name for DNS, URL for HTTP. The Total Logs means the number of different logs within 2 hours in CSTNET. The Matched Regex strings means the number of these top 50 websites captured in 2 hours. Total IP, URL and (IP, Domain) are the number of different cases in this 2 hours. The system needs to correct the 13,621 (IP, Domain) to obtain the 100\% correct ones.

\begin{table}
\caption{Dataset overview}
\centering
\begin{tabular}{lc}
\hline
\bfseries Overview & \bfseries Count\\
\hline
Total Logs & 73,627\\
Matched Regex strings & 47\\
Total IP & 11,360\\
Total URL & 50,267\\
Total (IP, Domain) & 13,621 \\
\hline
\end{tabular}
\label{tab:dataset_outline}
\end{table}

\subsection{Evaluation indicators}
The evaluation indicators is the recall and accuracy rate as calculated by Equation (\ref{eq:recall}), (\ref{eq:accuracy}). $N_{C\rightarrow C'}$ is the correct (IP, Domain) judged as correct and $N_{C\rightarrow I'}$ is the correct (IP, Domain) misjudge as incorrct. $N_{I\rightarrow C'}$ is the incorrect (IP, Domain) misjudged as correct and $N_{I\rightarrow I'}$ is the incorrect (IP, Domain) judged as incorrect.

Recall rate measures the rate of correct (IP, Domain) judged as correct out of the total correct (IP, Domain). Accuracy rate measures the rate of incorrect and correct (IP, Domain) which are judged correctly with respect to all the (IP, Domain).

\begin{equation}
TP=\frac{N_{C\rightarrow C'}}{N_{C\rightarrow C'}+N_{C\rightarrow I'}}
\label{eq:recall}
\end{equation}

\begin{equation}
ACC=\frac{N_{C\rightarrow C'}+N_{I\rightarrow I'}}{N_{C\rightarrow C'}+N_{C\rightarrow I'}+N_{I\rightarrow C'}+N_{I\rightarrow I'}}
\label{eq:accuracy}
\end{equation}

Given the heavy imbalance of the dataset (even if judging all (IP, Domain) are incorrect, the accuracy rate can achieve 98\%), add the precision rate, which calculated by Equation (\ref{eq:precision}), measures the rate of correct (IP, Domain) judged as correct out of the judged correct (IP, Domain).

\begin{equation}
P=\frac{N_{C\rightarrow C'}}{N_{I\rightarrow C'}+N_{C\rightarrow C'}}
\label{eq:precision}
\end{equation}

\subsection{Results}
As indicated in Table \ref{tab:expresult}, the system final results in the optimal SimHash threshold can achieve 94.30\% precision, 93.07\% recall and 99.78\% accuracy.

\begin{table}
\caption{Experiments results}
\centering
\begin{tabular}{lc}
\hline
\bfseries Results & \bfseries Count\\
\hline
Correct (IP, Domain) & 231\\
CDN server IP & 640\\
Non-Homepage URL & 10,347\\
Accuracy & 99.78\% \\
Recall & \bfseries 93.07\% \\
Precision & \bfseries 94.30\% \\
\hline
\end{tabular}
\label{tab:expresult}
\end{table}

\begin{figure}
\centering
\includegraphics[height=5cm]{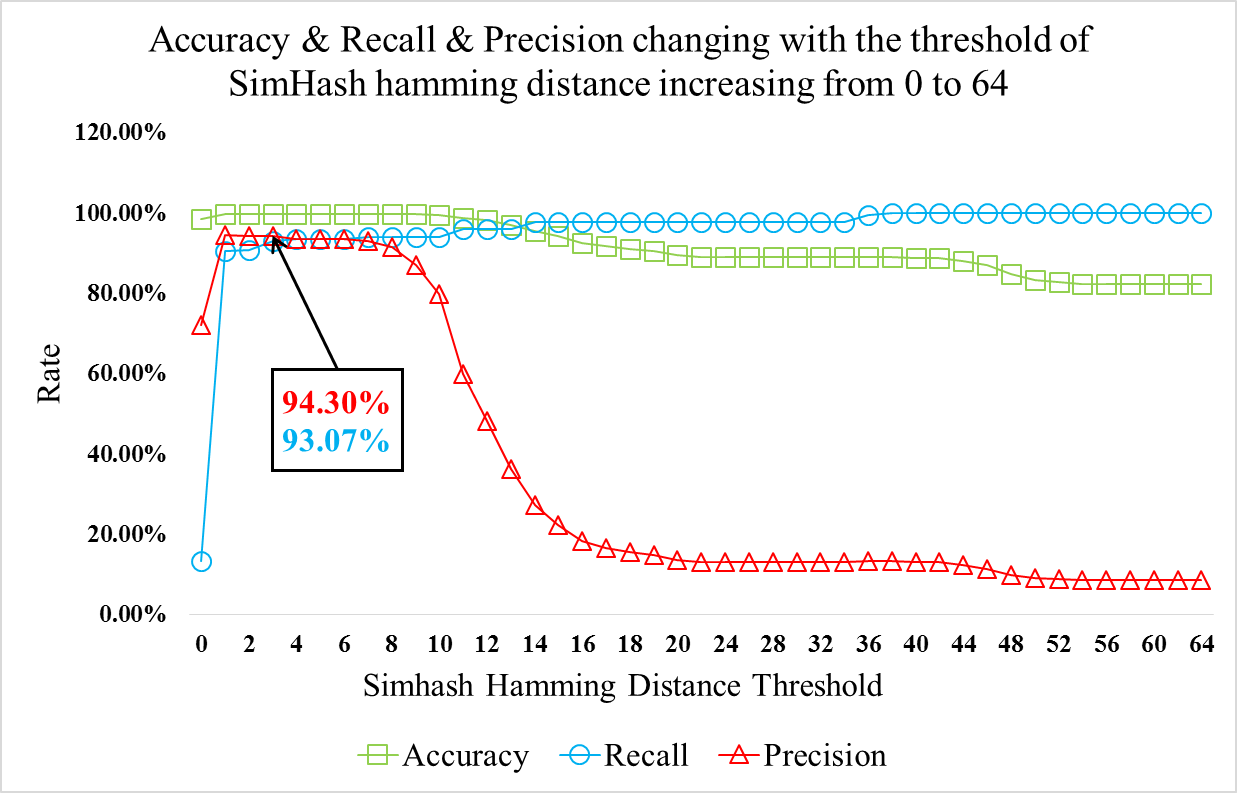}
\caption{The threshold of SimHash hamming distance versus Accuracy \& Recall \& Precision rate.}
\label{fig:expresult}
\end{figure}

For the different SimHash thresholds to the system, as Figure \ref{fig:expresult} displays, when the threshold is selected 3, the system achieves a good balance between the precision and recall rate.

\subsection{Online Testing}
We test SFCSD on the CSTNET, by using a server with the performance as Table \ref{tab:computerperformance} to evaluate the system's processing speed and the number of (IP, Domain) found. After a period of time running, SFCSD's online processing speed can achieve 8Gbps, due to the self-feedback mechanism, the system will gradually increase the automatic specific IP \& Domain strings. For the average of each Domain string, SFCSD can find 1000 possibly corresponded IP per day, and correct about 200.

\begin{table}
\caption{Standalone Performance}
\centering
\begin{tabular}{|l|l|}
\hline
Operation System & Red Hat Enterprise Linux 7.2 \\
\hline
CPU & Intel Xeon E5-4620 2.2GHz 8 Core \\
\hline
Memory & DDR3 256GB \\
\hline
Network Card & 10 Gigabit DPDK \\
\hline
Hard Dist & SATA 500GB \\
\hline
\end{tabular}
\label{tab:computerperformance}
\end{table}

\section{Discussion}
\label{sec:discussion}
\subsection{Limitation}
This system mainly has two limitations:

In order to meet the online processing speed under the condition of certain computing resources, SFCSD reduces the algorithm complexity of online key technology as much as possible, so the evasion method is very simple that is to change the CDN CNAME or homepage URL not to contain the feature strings. But considered the certain strings almost are the company's name to represent who is it, so the feature string can still work by tracking continuously.

The webpage fingerprint algorithm has a premise that the webpage content is the plaintext from HTTP, but for HTTPs, this method will be immediately invalid. Given this problem is aim to all the passive measurement, not just for SFCSD. It still can filter incorrect (IP, Domain) by the regex matching, CDN filter to reduce manual correction workload as much as possible.

\subsection{Future Work}
Future work will aim at above limitations and be divided into two aspects. One is to extract a few other webpage characteristics to assist the SimHash and improve the whole system's effects. Another is try to analyze some previous work about identifying the similar cipher text without decryption \cite{buyrukbilen2013secure} for HTTPS.

\section{Conclusion}
\label{sec:conclusion}

Currently, there are many malicious or non-malicious activities which lead to the incorrect DNS resource records, such as server configuration errors, DNS caching poisoning and DNS spoofing, etc.

This paper focuses on whether the DNS resource records can have direct access to the websites correctly, proposing a self-feedback correction system for DNS (SFCSD) based on active and passive measurement.

Through the possibility calculating method proposed as the theoretical basis of the self-feedback mechanism and all functions work together, it can find 1000 possibly corresponded IP per day for the average of each Domain string on the CSTNET.

By utilizing SSL, DNS and HTTP traffic together, filtering with 8 CDN and 8 URL features, verifying by SimHash, the passive correction can achieve 94.3\% precision and 93.07\% recall rate when the threshold of SimHash hamming distance sets 3 in the test dataset.

Combined with the active correction, under the above conditions, SFCSD can achieve 8Gbps processing line speed stand-alone and correct about 200 corresponded IP per day for the average of each Domain string.

\section*{Acknowledgment}
This work is supported by the National Key R\&D Program with No.2016YFB081304, National Natural Science Foundation of China (No. 61402464, No. 61402474). The corresponding author is xueqiang zou.

\bibliographystyle{IEEEtranS}
\bibliography{bare_conf}

\end{document}